\documentclass[conference]{IEEEtran}
\IEEEoverridecommandlockouts
\usepackage{cite}
\usepackage{amsmath,amssymb,amsfonts}
\usepackage{algorithmic}
\usepackage{graphicx}
\usepackage{textcomp}
\usepackage{xcolor}
\usepackage{comment}
\usepackage{orcidlink}
\usepackage{subfig}

\usepackage{graphicx}
\usepackage{xcolor}
\usepackage{url}
\usepackage{ulem}
\usepackage{booktabs}
\usepackage{makecell}
\usepackage{adjustbox}
\usepackage{enumitem}
\usepackage{wrapfig}
\usepackage{tikz}

\newcommand{\hyeran}[1]{\textcolor{black}{#1}}

\newcommand{\name}{\mbox{\textsc{AutoUVM}}}

\newcommand*\gcircled[1]{\tikz[baseline=(char.base)]{\node[circle, fill=darkgreen, inner sep=0.1ex, text=white] (char) {\scalebox{0.95}{#1}};}}
\newcommand*\bcircled[1]{\tikz[baseline=(char.base)]{\node[circle, fill=blue, inner sep=0.1ex, text=white] (char) {\scalebox{0.95}{#1}};}}
\definecolor{darkgreen}{rgb}{0.0, 0.5, 0.0}

\def\BibTeX{{\rm B\kern-.05em{\sc i\kern-.025em b}\kern-.08em
    T\kern-.1667em\lower.7ex\hbox{E}\kern-.125emX}}
\begin{document}

\bstctlcite{BSTcontrol}

\title{\name{}: Automated Prefetching Framework for LLMs under UVM Oversubscription
}

\author{\IEEEauthorblockN{
    Mao Lin\IEEEauthorrefmark{1}\orcidlink{0000-0002-1460-0766},
    Hui Feng\IEEEauthorrefmark{1},
    Xianzhong Ding\IEEEauthorrefmark{1},
    Guilherme Cox\IEEEauthorrefmark{2},
    Qian Wang\IEEEauthorrefmark{1},
    Hyeran Jeon\IEEEauthorrefmark{1}\orcidlink{0000-0002-1767-8198}
}\\
  \IEEEauthorblockA{\IEEEauthorrefmark{1}University of California, Merced
  \quad
  \IEEEauthorrefmark{2}NVIDIA}
}

\maketitle

\begin{abstract}
Large language models (LLMs) increasingly exceed the memory capacity of commodity GPUs, making memory oversubscription common in practical deployments. NVIDIA Unified Virtual Memory (UVM) provides transparent access to host memory, but its page-fault–driven migrations introduce severe performance overhead. While UVM exposes primitives (e.g., prefetching and placement hints) to mitigate these costs, they require low-level CUDA modifications, limiting their applicability for most LLM users. Meanwhile, existing UVM optimizations operate at coarse managed-object granularity and fail to capture deep learning frameworks’ internal tensor-level memory behavior, leading to excessive data movement and CPU–GPU interconnect bottlenecks.

We propose \name{}, an automated, framework-aware UVM prefetching system for efficient LLM execution under memory oversubscription. \name{} bridges the semantic gap between deep learning frameworks and UVM by exposing tensor-level access information and enabling policy-driven prefetching at fine granularity. Implemented as a transparent extension, \name{} requires no changes to model code and dynamically adapts to runtime memory pressure.
We instantiate \name{} with a roofline-inspired policy to identify performance-critical data transfers.
Across ten LLMs, \name{} achieves an average $3.1\times$ speedup over baseline UVM and consistently surpasses the best-performing prior UVM prefetcher by $1.9\times$, with improvements of up to $4.7\times$ over object-level prefetchers, while significantly reducing page faults.
\end{abstract}


\section{Introduction}

Memory demand continues to grow rapidly in modern data-intensive workloads, especially for large language model (LLM) training and serving, making GPU memory capacity a critical bottleneck. To mitigate memory pressure, prior work has explored multi-GPU parallelization~\cite{ zhao2023pytorch}, host-memory offloading~\cite{lin2023drgpum}, activation recomputation~\cite{pmlr-v162-patil22b}, and memory compression or quantization~\cite{young2019enabling}. However, these approaches often require workload-specific tuning, specialized hardware support, or remain fundamentally limited by GPU memory capacity.

To alleviate limited GPU memory capacity, 
NVIDIA’s Unified Virtual Memory (UVM) can be used. UVM unifies CPU and GPU address spaces and enables on-demand data migration between host and device memory. With UVM, applications can allocate memory beyond the GPU’s physical capacity, enabling memory oversubscription with minimal programmer intervention. With 
prefetch APIs, such as \texttt{cudaMemPrefetchAsync} and \texttt{cudaMemAdvise}, 
developers can explicitly control and optimize the migration and placement of UVM-managed data objects. However, manual UVM management requires substantial engineering effort, including source-code modifications and workload-specific expertise. 
In contrast, UVM's autonomous memory-management mechanisms, including its tree-based prefetch engine and follow-up variants~\cite{lin2025forest, suv2024, Ganguly2019inter}, exploit coarse-grained locality to prefetch and evict data without programmer intervention.

Existing programmer-agnostic UVM prefetching approaches operate at the granularity of UVM-managed objects by leveraging the UVM driver's object-level access monitoring capability
~\cite{lin2025forest, suv2024, Ganguly2019inter, lin2023drgpum}. However, modern deep learning (DL) frameworks introduce a semantic gap between framework-level tensor management and the UVM driver. Frameworks such as PyTorch and TensorFlow employ internal 
allocators (e.g., the PyTorch Caching Allocator (PCA))~\cite{pytorch-pca, lin2025understanding}, that reserve large monolithic memory pools and 
sub-allocate them to many tensors. These sub-allocations are invisible to the UVM driver; from its perspective, tensor accesses appear as reuse within a few large managed objects. Consequently, the UVM driver cannot observe tensor-level memory activities, preventing accurate tensor-level prefetching and limiting the effectiveness of autonomous UVM optimizations. 
This mismatch fundamentally limits the effectiveness of prior UVM optimizations, not only prefetching, but also 
page placement, proactive eviction, and offloading.

To address these limitations, a practical UVM solution must (1) recover tensor-level access information hidden by DL frameworks, (2) identify which tensor migrations are performance-critical, and (3) dynamically adapt decisions as memory pressure and eviction behavior evolve during execution.

\textbf{In this work, we propose \name{}, an automated, framework-semantic-aware UVM prefetching framework for oversubscribed LLM workloads.} 
Because tensor-level memory activities are hidden behind framework-managed memory pools, the UVM driver cannot observe which tensors are accessed by individual kernels at runtime. To recover this missing information without modifying model code,
\name{} performs a single offline profiling pass to recover kernel--tensor access relationships. 
While profile-guided prefetching with runtime monitoring has been extensively explored~\cite{peng2020capuchin, yuan2026cost, lin2026pasta, lin2026hybridgen}, \name{} is the first to bring it to the UVM path and enable automated tensor-level prefetching without modifying model code. During execution, \name{} automatically issues policy-driven, fine-grained tensor-level prefetches. 
Since LLM kernels exhibit heterogeneous sensitivity to CPU--GPU migration latency, blindly prefetching all tensors can waste CPU--GPU interconnect (CGI) bandwidth and increase GPU-memory pressure.
To selectively target migrations that are likely to improve performance, \name{} decouples framework-semantic extraction from prefetch-policy design, enabling different policies to be deployed across workloads and hardware platforms.
In this paper, we instantiate \name{} with a roofline-inspired policy that distinguishes CGI-bound, GPU-memory-bound, and compute-bound kernels.
Furthermore, because oversubscription-induced evictions can alter runtime behavior, \name{} combines offline profiles with a lightweight online executor that continuously adapts prefetching decisions as execution conditions evolve.

The key insight behind \name{} is that effective UVM management requires both tensor-level visibility and selective prefetching. Prior approaches operate without tensor semantics, which is fundamentally misaligned with tensor-level memory management in DL frameworks and often results in excessive data migration. By combining tensor-aware profiling with roofline-inspired analysis, \name{} reduces unnecessary migrations, mitigates CGI bottlenecks, reduces page faults, and enables predictable, performance-aware UVM behavior for memory-intensive workloads.

Our contributions are as follows:

\begin{enumerate}[label=\textbullet, leftmargin=*, labelindent=2pt]
\item We reveal the limitations of monolithic object-level UVM prefetching in modern DL frameworks and introduce \name{}, the first automated, policy-driven UVM prefetching system with tensor-level visibility.

\item We develop a profile-driven analysis framework that performs a single offline pass to trace kernel--tensor accesses, generate stable tensor identifiers, and infer tensor affinity, enabling framework-aware prefetch planning without modifying model code. We also design a lightweight Online Executor that autonomously issues and adapts tensor-level prefetching decisions using a lightweight runtime monitor.

\item We instantiate \name{} with a roofline-inspired policy to identify performance-critical data transfers. Across ten LLMs, \name{} achieves an average $3.1\times$ speedup over baseline UVM, outperforms the best-performing prior prefetcher (DeepUM) by $1.9\times$, and improves over object-level prefetchers by up to $4.7\times$, while significantly reducing page faults.
\end{enumerate}

\section{Background}
\begin{figure}[t]
    \centering
    \includegraphics[width=0.80\linewidth]{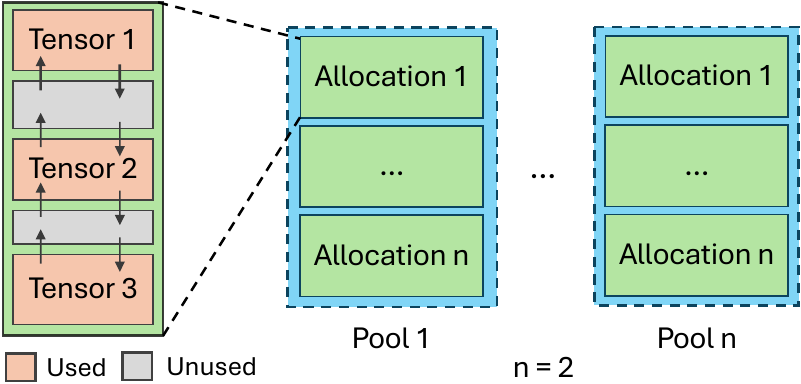}
    \caption{Memory Management in the PyTorch Framework}
    \label{fig:pca}
    \vspace{-10pt}
\end{figure}

\subsection{Unified Virtual Memory}

Modern heterogeneous systems with discrete CPUs and GPUs maintain separate physical memories connected through CPU-GPU interconnect (e.g., PCIe, NVLink), traditionally requiring explicit allocation and copy operations. To simplify programming, NVIDIA introduced Unified Virtual Memory (UVM) in CUDA~6.0~\cite{uvm-cuda6}, enabling CPUs and GPUs to share data through a unified pointer. Pascal architecture~\cite{pascal-architecture} extended UVM with on-demand page migration and memory oversubscription. With \texttt{cudaMallocManaged}, data pages migrate to GPU memory on first access; if a nonresident page is referenced, a page fault triggers the UVM driver and GPU Memory Management Unit (GMMU) to allocate device space and copy the page from host to device.

Although UVM removes explicit data transfers, page-fault handling introduces substantial latency. NVIDIA provides advisory APIs such as \texttt{cudaMemPrefetchAsync} for proactive migration and \texttt{cudaMemAdvise} for preferred placement. 

UVM offers a flat virtual address space shared by CPU and GPU and supports \textit{memory oversubscription}, where CPU memory backs allocations exceeding GPU capacity. When GPU memory is full, the UVM driver evicts least recently used (LRU) pages to host memory. Oversubscription severity is expressed by the \textit{oversubscription factor}, the ratio of allocated UVM memory to GPU capacity~\cite{uvm-oversubscription}. A factor $\leq 1.0$ indicates no oversubscription, while $>1.0$ requires frequent evictions; for instance, a factor of 1.5 on an 80~GB GPU corresponds to 120~GB of managed memory, causing continuous page movement between CPU and GPU.

\subsection{Memory Management of Modern DL 
Frameworks}
DL 
workloads would suffer substantial overhead if each tensor is individually managed via costly \texttt{cudaMalloc} and \texttt{cudaFree} APIs, given the increasing DL model sizes. To avoid this, PyTorch~\cite{paszke2019pytorch} employs the PyTorch Caching Allocator (PCA)~\cite{pytorch-pca}, which uses a pooling strategy to amortize allocation costs. Instead of allocating memory for each tensor, PCA obtains large contiguous blocks via \texttt{cudaMalloc} and subdivides them for tensor use. Unused segments are retained in pools for reuse, and freed tensors return their space to the pool rather than triggering \texttt{cudaFree}. After a brief warm-up phase, most tensor allocations are served from cached blocks.

Figure~\ref{fig:pca} illustrates this design. PyTorch maintains separate large and small pools, each consisting of multiple pre-allocated blocks. Within each block, many tensors coexist and are tracked by a framework-level metadata structure (e.g., a doubly linked list) that records allocation boundaries and reuse opportunities. These behaviors occur entirely inside the framework and remain invisible to the GPU runtime, which sees only a small number of large allocations. Consequently, low-level allocators and profilers have limited visibility into the fine-grained tensor reuse and pooling dynamics managed by PCA.

\section{Motivation}\label{sec:motiv} 
\subsection{High Memory Demand Calls for UVM}

\begin{figure}[t]
    \centering
    \includegraphics[width=0.9\linewidth]{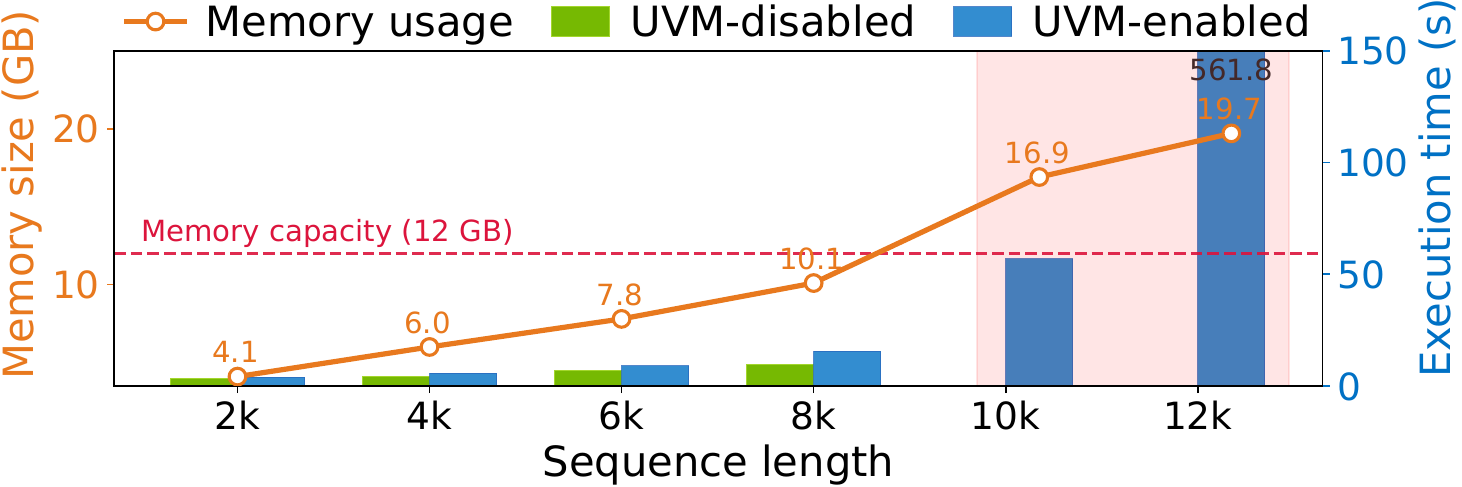}
    \caption{
    Qwen1.5 performance with and without UVM and memory usage across varying sequence lengths.}
    \label{fig:uvm_cmp_perf}
\end{figure}
With its support for memory oversubscription, UVM is the only developer-agnostic mechanism on NVIDIA GPUs that enables large workloads to run on devices with limited memory capacity, despite its inherent overheads. As the backend engine of emerging memory-management interfaces, such as Heterogeneous Memory Management (HMM)~\cite{linux-hmm} and zero-copy transfer in Grace–Hopper architectures~\cite{grace-hopper-whitepaper}, UVM is 
one of the most essential technologies to provide flexible and transparent CPU--GPU data movement.

Figure~\ref{fig:uvm_cmp_perf} shows this benefit for Qwen1.5 inference on an RTX~3060. Without UVM, PyTorch fails once sequence length exceeds 10K, resulting in an out-of-memory (OOM) error that would normally require multiple GPUs. With UVM enabled, the same GPU can execute significantly longer sequences, even when memory usage exceeds its 12~GB device capacity. As LLM sizes continue to grow and high-end GPUs remain expensive, UVM offers a practical path for enabling large-model workloads on commodity hardware.

However, UVM performance can degrade sharply under oversubscription due to the substantial cost of page-fault-driven on-demand migrations. Designing an effective prefetching strategy is therefore essential to fully unlocking UVM’s potential.

\subsection{Limitations of Existing UVM Prefetching}
\label{sec:object-level-prefetch}

Existing UVM prefetching techniques lack visibility into framework-level tensor-management semantics, making prefetching decisions using only UVM-level information~\cite{Ganguly2019inter, suv2024, lin2025forest, jung2023deepum}. Object-level approaches~\cite{suv2024, lin2025forest} further operate at the granularity of \texttt{cudaMallocManaged} allocations, assuming that all data within a managed object exhibits similar access behavior. However, this assumption breaks down for large LLM workloads running on modern deep learning frameworks such as PyTorch. A single managed allocation often contains many tensors with heterogeneous access patterns and lifetimes. Without visibility into tensor-level behavior, existing prefetchers fail to capture the fine-grained memory needs of individual kernels, resulting in inefficient migration.

\begin{figure}[t]
    \centering
    \includegraphics[width=0.8\linewidth]{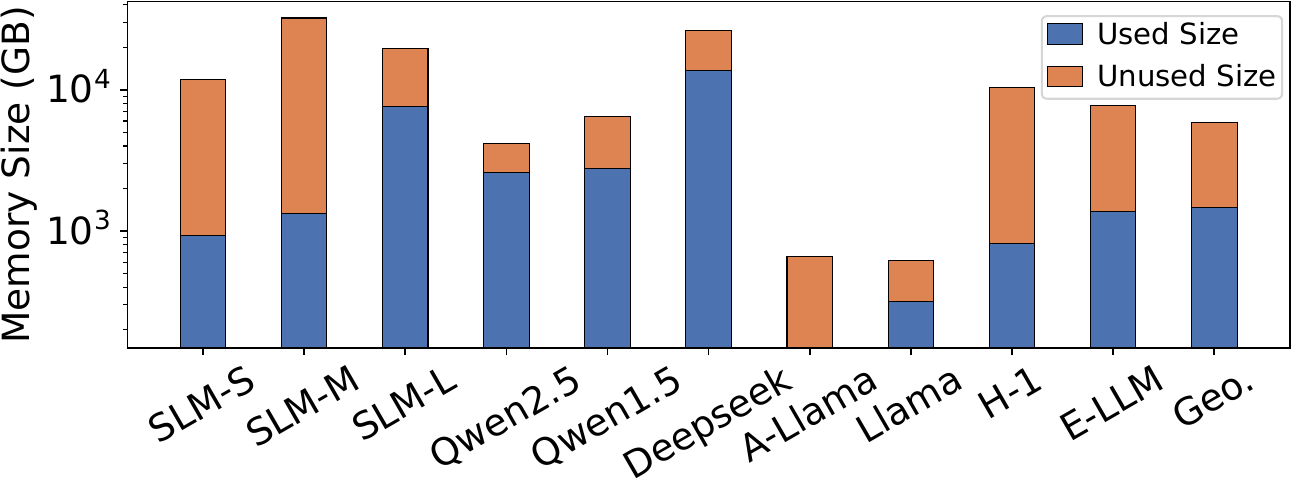}
    \caption{Breakdown of used and unused prefetch sizes for object-level UVM prefetching in PyTorch, when the prefetcher is unaware of PyTorch’s internal memory management (log scale).}
    \vspace{-10pt}
    \label{fig:unused_memory}
\end{figure}

To quantify this inefficiency, we implement a naive object-level strategy that prefetches entire managed objects immediately before each kernel launch. Figure~\ref{fig:unused_memory} shows the breakdown of used versus unused prefetched data across several LLM models (detailed in Table~\ref{tab:models}). On average, useless prefetch is 6$\times$ (up to 23$\times$) more than useful prefetch. Such useless prefetches waste both CGI bandwidth and scarce GPU memory capacity, illustrating that coarse-grained, application-agnostic prefetching is ill-suited to the fine-grained memory access patterns of LLM workloads.

\subsection{Heterogeneous Migration Sensitivity Across Kernels}
\label{sec:roofline}
\begin{figure}[t]
    \centering
    \includegraphics[width=0.95\linewidth]{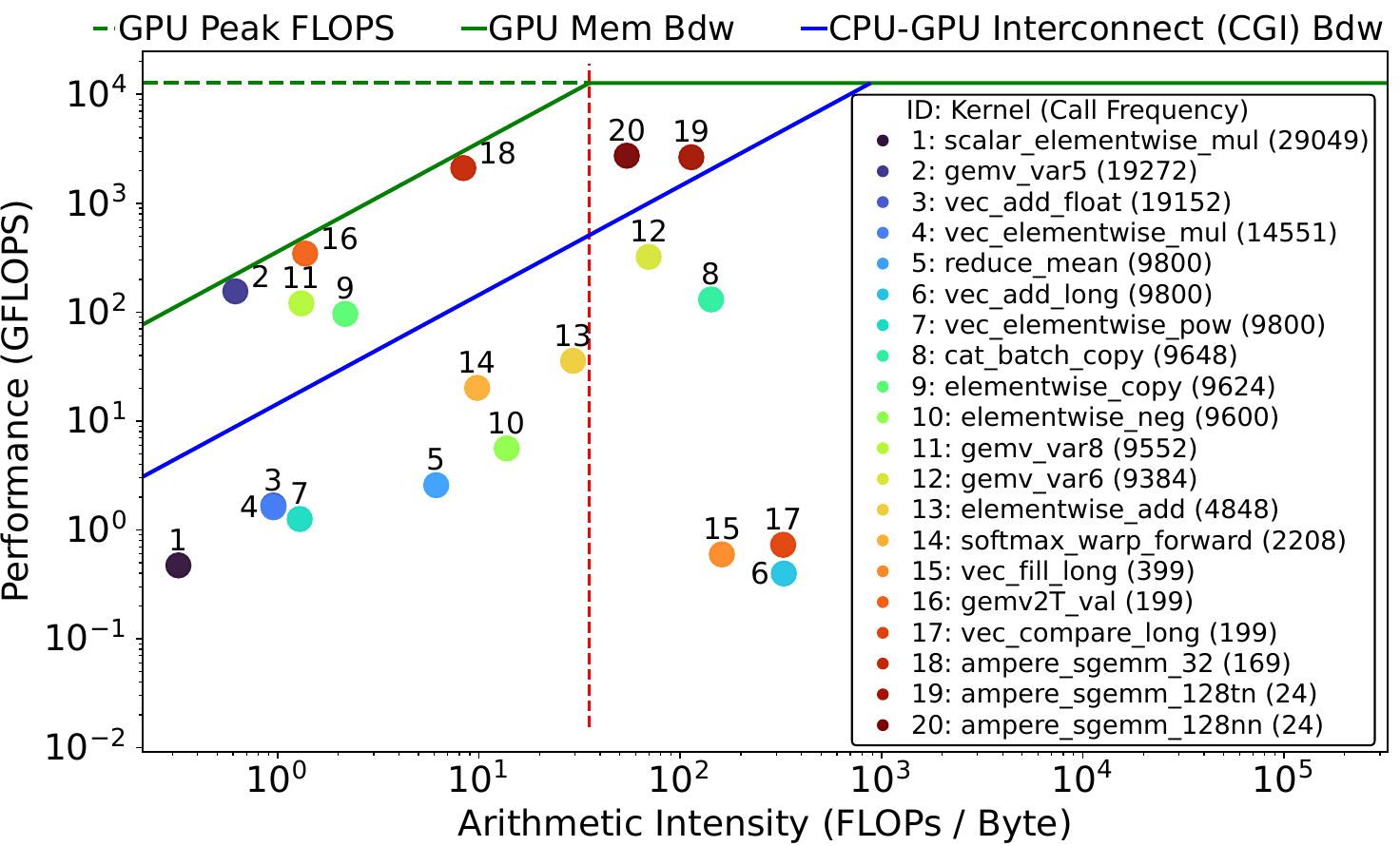}
    \caption{Roofline visualization of Qwen1.5 running on NVIDIA RTX 3060. The figure highlights the top 20 kernels, which dominate performance and together contribute to more than 90\% of overall runtime.} 
    \label{fig:roofline}
    \vspace{-10pt}
\end{figure}

Different kernels exhibit different sensitivity to CPU--GPU migration latency. Memory-bound kernels rely heavily on fast data access, whereas compute-bound kernels are typically more tolerant. LLM workloads contain many such kernel types, often invoked thousands of times during inference.

Figure~\ref{fig:roofline} shows this heterogeneity using a multi-level roofline analysis of Qwen1.5 on an NVIDIA RTX~3060. The plot includes the GPU compute roof, GPU memory-bandwidth roof, and the significantly lower CPU--GPU interconnect (CGI) roof that reflects transfer limits over PCIe. Elementwise kernels cluster in the low-intensity memory-bound region, while GEMV/GEMM kernels approach the compute roof. Several operators fall near or below the CGI roofline, indicating substantial throughput degradation once tensor pages migrate between CPU and GPU memory via UVM. This observation highlights that blindly prefetching all tensors is inefficient; prefetching should instead prioritize kernels most sensitive to interconnect bottlenecks.
Note that this finding applies even to the emerging GPUs that have significantly higher CGI bandwidth, because GPU memory bandwidth (e.g., 4 TBps in Hopper GPU) is typically significantly higher than the CGI's (e.g., 900 GBps in NVlink).

\section{Design of \name{}}

\subsection{Design Goals}
Motivated by the observations in Section~\ref{sec:motiv}, \name{} aims to bridge the semantic gap between DL frameworks and the UVM driver to enable efficient execution under memory oversubscription.
The design of \name{} is guided by four goals:
\textit{(1) Tensor awareness.} Prefetching should operate at tensor granularity rather than coarse \texttt{cudaMallocManaged} objects.
\textit{(2) Policy-driven selectivity.} Prefetching should target kernels most sensitive to CPU--GPU migration overhead.
\textit{(3) Runtime adaptivity.} Prefetch decisions should adapt to changing memory pressure and eviction behavior.
\textit{(4) Framework transparency.} The system should require no model-code modifications and integrate seamlessly with DL frameworks such as PyTorch.

\subsection{Key Ideas}
The core insight behind \name{} is that effective UVM prefetching requires both tensor-level visibility and selective migration. Rather than assuming all kernels benefit equally from proactive migration, \name{} applies policy-driven prefetching based on workload and runtime characteristics. In this work, we instantiate \name{} with a roofline-inspired policy that prioritizes kernels bottlenecked by CPU--GPU interconnect (CGI) transfers while avoiding unnecessary migration for compute-bound and GPU-memory-bound kernels, as discussed in Section~\ref{sec:roofline}.

Prefetching at the \texttt{cudaMallocManaged} object level is too coarse because framework allocators pack many unrelated tensors into large shared regions, causing conventional prefetching to migrate substantial unused data. \name{} addresses this by identifying the exact tensors each kernel accesses and migrating only those. To ensure these profiles remain valid across runs, \name{} assigns each tensor a stable identity derived from its creation-site call stack and a global allocation counter, making it robust to address randomization by the OS or framework. These ideas collectively form the foundation of \name{}'s design.

\subsection{System Overview}

\begin{figure}[t]
    \centering
    \includegraphics[width=0.95\linewidth]{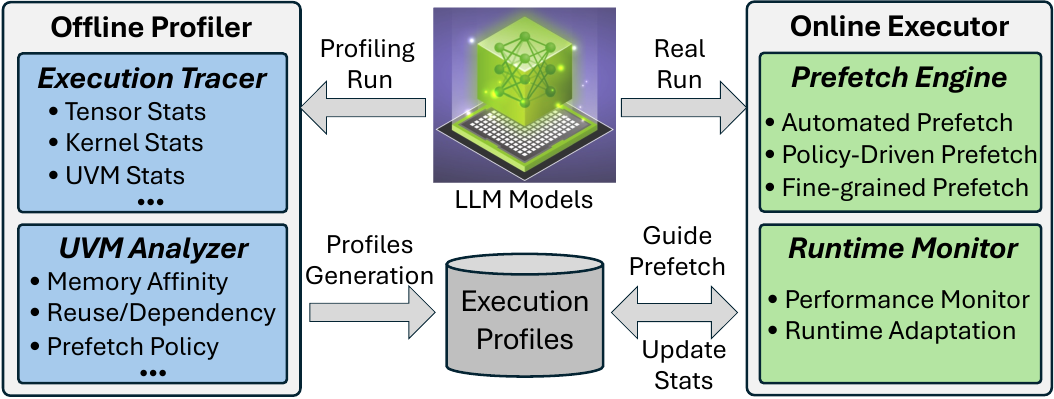}
    \caption{Design of \name{}.}
    \label{fig:design}
\end{figure}

Figure~\ref{fig:design} presents the architecture of \name{}, which consists of an \textit{Offline Profiler}, an \textit{Online Executor}, and an \textit{Execution Profile Database}. The Offline Profiler observes a single profiling run to capture kernel--tensor access patterns, collect runtime statistics, and infer tensor affinity across consecutive kernels. The resulting execution profiles encode kernel behaviors, kernel--tensor mappings, and metadata required to identify tensors across executions.

During inference, the Online Executor loads the execution profiles and issues policy-driven prefetch operations before kernel launches. As oversubscription pressure changes during execution, the Online Executor continuously monitors runtime behavior and dynamically adjusts prefetching decisions. This adaptive mechanism ensures that prefetching is applied only when beneficial, even when runtime evictions cause execution behavior to deviate from the offline profile.

\subsection{Offline Profiler}
\subsubsection{Execution Tracer}
Profiling-based memory optimization for LLMs faces two challenges that prevent naive reuse of profiling results:
1) \textit{Dynamic computation graphs.}  
Frameworks like PyTorch construct graphs eagerly, causing operator sequences to differ across runs, making static prefetch schedules unreliable.
2) \textit{Address randomization.}  
Tensor addresses change between profiling and execution, invalidating pointer-based profiles. To address these issues, we employ two techniques:

\textit{Kernel--Tensor Access Tracing.}
Instead of relying on the computation graph, the Execution Tracer records \textit{kernel--tensor access pairs}. During execution, the Online Executor simply prefetches the tensors needed by the next kernel.

\textit{Two-Level Tensor Identifier.}
To track tensors across runs, \name{} assigns each tensor a stable identity derived from its creation-site call stack and a monotonic allocation counter. PyTorch’s caching allocator is instrumented with a lightweight hook to record this information. If a mismatch occurs at runtime, \name{} disables prefetch for the affected tensors to ensure correctness, though such cases are rare for fixed-architecture LLMs.

\subsubsection{UVM Analyzer}
\label{roofline_analyzer}

The UVM Analyzer analyzes kernel behavior and runtime statistics to guide policy-driven prefetching. In this work, we instantiate the analyzer with a roofline-inspired policy that classifies kernels into GPU-compute-bound, GPU-memory-bound, and CGI-bound categories. Since kernels launched from the same call stack and configuration typically exhibit similar behavior, \name{} applies a unified prefetch strategy across such instances.

In addition, the Analyzer performs \textit{look-ahead tensor affinity analysis} to identify tensors reused across consecutive kernels. Tensors with high short-term reuse are marked for GPU residency using \texttt{cudaMemAdvise}, reducing migration overhead during oversubscription.

\subsubsection{Execution Profiles}
The Offline Profiler consolidates collected information into compact execution profiles that map kernels to accessed tensors. Kernels are identified using launch-site call-stack hashes and launch counters, while tensors use creation-site call-stack hashes and allocation counters. The profiles additionally store runtime metadata, policy information, and tensor-affinity hints used to guide subsequent prefetching and pinning decisions during real execution.

\subsection{Online Executor}

\subsubsection{Prefetch Engine}

Unlike prior UVM techniques that operate at the granularity of entire managed objects, the Prefetch Engine performs fine-grained tensor-level prefetching, migrating only the tensors required by upcoming kernels. This selective approach reduces unnecessary data movement and improves GPU memory utilization under oversubscription. 
Prefetch decisions are guided by the selected policy. In our roofline-inspired policy, kernels bottlenecked by CPU--GPU interconnect (CGI) transfers trigger proactive prefetching to hide migration latency, while compute-bound and GPU-memory-bound kernels avoid unnecessary migration to conserve bandwidth and reduce memory pressure.
For tensors reused across consecutive kernels, the Prefetch Engine uses \texttt{cudaMemAdvise} to pin them on the GPU during their reuse window. The runtime mechanism relies only on lightweight hooks on kernel launches and tensor allocations, keeping runtime overhead minimal.

\subsubsection{Runtime Monitor}

Under oversubscription, runtime evictions may alter kernel behavior and reduce the effectiveness of static prefetch decisions. To remain adaptive, the Runtime Monitor continuously tracks kernel performance during execution and updates policy decisions accordingly. In our roofline-inspired policy, the monitor dynamically re-evaluates kernel classifications based on updated runtime measurements (e.g., kernel time). When a classification shift is detected, the Prefetch Engine immediately adjusts its migration decisions to match current memory pressure. The Runtime Monitor introduces only lightweight timestamping overhead, while policy updates execute in parallel with subsequent kernels, resulting in negligible runtime cost.


\begin{figure}[t]
    \centering
    \includegraphics[width=0.85\linewidth]{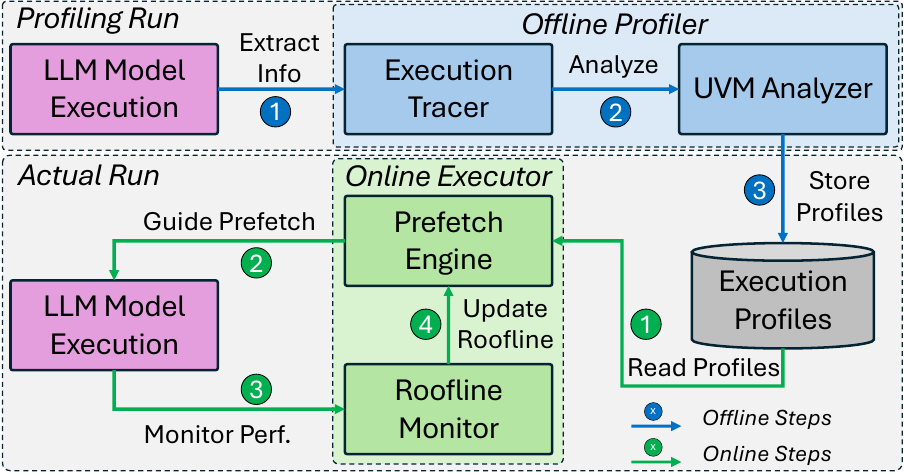}
    \caption{Workflow of \name{}.}
    \vspace{-5pt}
    \label{fig:workflow}
\end{figure}

\subsection{Workflow}

Figure~\ref{fig:workflow} shows the two-phase workflow of \name{}. During the profiling phase, the Execution Tracer extracts kernel--tensor access relationships from an LLM execution (\bcircled{1}), and the UVM Analyzer analyzes runtime statistics to generate execution profiles (\bcircled{2}). The generated profiles are then stored in the Profile Database for reuse across executions (\bcircled{3}).
During online execution, the Online Executor loads the execution profiles at startup (\gcircled{1}) and uses them to guide tensor prefetching before kernel launches (\gcircled{2}). As execution progresses, the Runtime Monitor continuously tracks kernel performance (\gcircled{3}) and updates runtime roofline information under changing memory pressure. The Prefetch Engine then dynamically adjusts prefetching decisions based on the updated runtime behavior (\gcircled{4}).

\subsection{System Implementation}
Our implementation consists of a UVM profiler and a prefetch engine. \name{} is integrated into PyTorch as a transparent extension that requires no changes to model code. We use tensor-creation hooks~\cite{tensor-hook} in PyTorch’s CUDA caching allocator to record call stacks, tensor sizes, and allocation counters. NVIDIA NVBit~\cite{villa2019nvbit} is used to count FLOP instructions and memory loads.
NVIDIA Compute Sanitizer APIs~\cite{compute-santizer-api} are used to inject prefetch and affinity primitives with minimal overhead. To enable UVM semantics, we modify PyTorch’s caching allocator to use \texttt{cudaMallocManaged} in place of \texttt{cudaMalloc}, ensuring full UVM compatibility.
We evaluate the power and energy impact of \name{} using a lightweight Python measurement tool built on NVML that samples GPU power sensors and accumulates total energy during execution.


\section{Evaluation}
\subsection{Methodology}

\begin{table}[t]
\caption{Hardware and Software Environment.}
\label{tab:platform}
\centering
\scriptsize
\begin{adjustbox}{width=0.45\textwidth}
\begin{tabular}{|c|c|c|c|c|c|c|c|}
\hline
CPU & GPU & System & \makecell{System\\Memory} & \makecell{GPU\\Driver} & \makecell{GPU \\Toolkit} \\
\hline
\makecell{AMD Ryzen 7\\5800X} & \makecell{NVIDIA GeForce\\RTX 3060} & Linux 6.11 & 32 GB & 560.28.03 & CUDA 12.1 \\
\hline
\makecell{Intel Xeon\\Gold 5320} & \makecell{NVIDIA A100} & Linux 6.16 & 128 GB & 580.65.06 & CUDA 12.1 \\
\hline
\makecell{AMD\\PYC 9454} & \makecell{NVIDIA\\H100 NVL} & Linux 5.15 & 560 GB & 560.35.05 & CUDA 12.4 \\
\hline
\end{tabular}
\end{adjustbox}
\end{table}

\begin{table}[t]
\caption{Evaluated LLM models.}
\label{tab:models}
\centering
\scriptsize
\begin{adjustbox}{width=0.42\textwidth}
\begin{tabular}{|c|c|c|c|c|c|c|c|}
\hline
LLM Models & \makecell{Transformer\\Layers} &  \makecell{Sequence\\Length} & \makecell{Memory\\Footprint} & Abbr.\\
\hline
\hline
SmolLM2-S~\cite{allal2025smollm2smolgoesbig} & 30 & 4.0k & 3.6GB & SLM-S\\
\hline
SmolLM2-M~\cite{allal2025smollm2smolgoesbig} & 32 & 4.0k & 4.8GB & SLM-M\\
\hline
SmolLM2-L~\cite{allal2025smollm2smolgoesbig} & 24 & 4.0k & 11.4GB & SLM-L\\
\hline
Qwen2.5~\cite{qwen2.5}  & 24 & 4.0k & 6.9GB & Qwen2.5\\
\hline
Qwen1.5~\cite{qwen1.5} & 24 & 4.0k & 11.2GB & Qwen1.5\\
\hline
Deepseek-coder~\cite{deepseek-coder}  & 24 & 4.0k & 10.2GB & Deepseek\\
\hline
AMD-Llama~\cite{amd-llama}  & 12 & 1.9k & 1.4GB & A-Llama\\
\hline
Llama~\cite{miao2023specinfer}  & 12 & 1.9k & 1.5GB & Llama\\
\hline
Helium-1~\cite{helium-1-2b}  & 28 & 3.9k & 10.8GB & H-1\\
\hline
EuroLLM~\cite{martins2025eurollm}  & 24 & 3.9k & 11GB & E-LLM\\
\hline
\end{tabular}
\end{adjustbox}
\vspace{-5pt}
\end{table}

We evaluated \name{} on three GPU platforms, as summarized in Table~\ref{tab:platform}.
We evaluate \name{} across 10 widely used LLMs from Hugging Face, listed in Table~\ref{tab:models}. To precisely control the degree of UVM oversubscription, we constrain the effective GPU memory capacity by allocating a fixed amount of device memory in a separate helper process, following standard practice in prior work~\cite{ganguly2020adaptive, lin2025understanding}.
We measure the number of page faults by adding a counter to the NVIDIA UVM driver module~\cite{nvidia-open-kernle}.

\subsubsection{Compared solutions}
We compare \name{} against a baseline and three UVM prefetching approaches: SUV and Forest, which operate at the managed-object level, and DeepUM, which operates at the UM-block level.

\begin{enumerate}[label=\textbullet,leftmargin=*, labelindent=0pt]
\item \textbf{Baseline} uses UVM without explicit prefetching. When a kernel accesses data not resident on the GPU, a page fault triggers on-demand migration, reflecting default UVM behavior.

\item \textbf{SUV}~\cite{suv2024} optimizes UVM at the object level by estimating each managed object's access density (GPU accesses per byte) via static analysis and pinning the object to host or device memory accordingly. We compute access density from memory-access traces collected by \name{}'s profiler.

\item \textbf{Forest}~\cite{lin2025forest} 
uses hardware counters to detect per-object access patterns and adjusts prefetch granularity. We emulate Forest using offline access-pattern analysis and vary granularity from 256~KB to 4~MB, matching its original setup. Because our pattern detection is offline, it avoids Forest’s online detection overhead.

\item \textbf{DeepUM}~\cite{jung2023deepum} uses correlation-based prefetching to predict future page accesses from historical fault patterns. 
Though effective for stable, repetitive access behavior, it operates at the coarse UM-block level, lacks framework-level semantics, and aggressively prefetches all correlated pages regardless of whether the movement is performance-critical.

\end{enumerate}

\begin{figure}[t]
    \centering
    \includegraphics[width=0.9\linewidth]{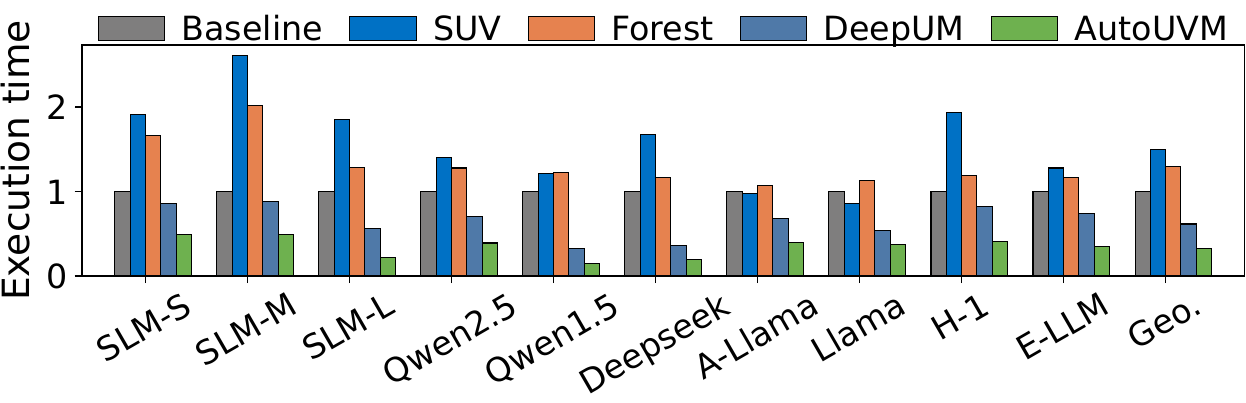}
    \caption{Execution time normalized to Baseline under memory oversubscription of 2.0 (lower is better).}
    \label{fig:overall_perf}
\end{figure}

\begin{figure}[t]
    \centering
    \includegraphics[width=0.9\linewidth]{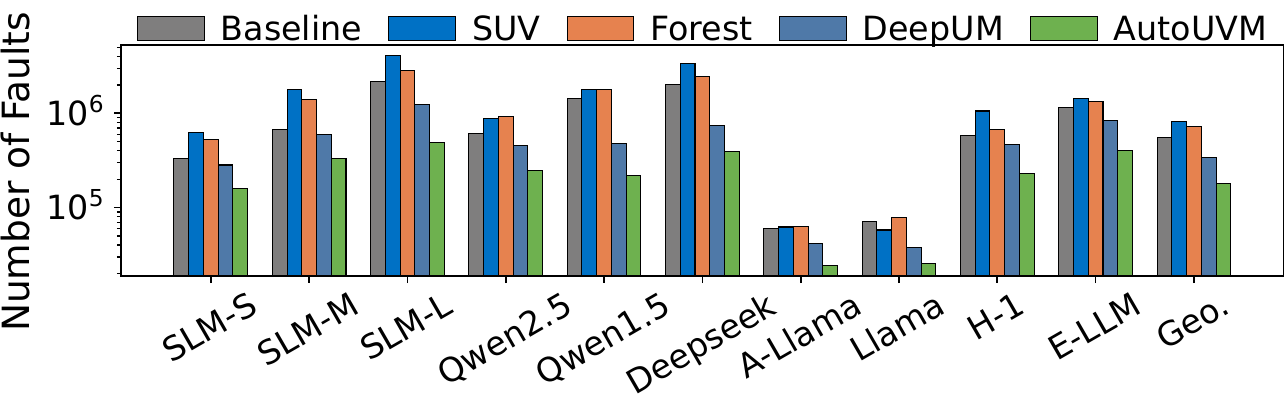}
    \caption{Number of page faults under memory oversubscription of 2.0 (lower is better).}
    \vspace{-10 pt}
    \label{fig:count_page_faults}
\end{figure}

\subsection{Overall Performance}

Figure~\ref{fig:overall_perf} reports execution time normalized to the Baseline under an oversubscription factor of 2.0 (lower is better). \name{} achieves average speedups of $3.1\times$, $4.7\times$, $4.1\times$, and $1.9\times$ over Baseline, SUV, Forest, and DeepUM, respectively. Both object-level approaches (SUV and Forest) often perform worse than the Baseline, showing that tensor-unaware prefetching is ineffective for LLM workloads on modern deep-learning frameworks. As discussed in Section~\ref{sec:object-level-prefetch}, object-level prefetching migrates entire UVM objects—including many unused tensors—which wastes CGI bandwidth, increases GPU-memory pressure, and triggers severe page thrashing. Between the two, Forest outperforms SUV by reducing prefetch granularity for irregular access patterns, limiting unnecessary migration.
DeepUM is the best-performing baseline, as its finer UM-block granularity and UVM-aware correlation prefetching consistently outperform the Baseline. However, it lacks framework-level semantics and prefetches correlated blocks regardless of whether the transfer is performance-critical, resulting in more data movement than \name{}.
By prefetching only the tensors each kernel actually accesses, \name{} migrates far fewer bytes and achieves the best results.

\subsection{Page Fault}
We measure the total number of page faults, as shown in Figure~\ref{fig:count_page_faults}. A strong correlation emerges between execution time and the number of page faults across all models and methods. \name{} issues the fewest page faults among the evaluated solutions, due to its policy-driven, tensor-level prefetching strategy, which allocates limited \hyeran{CGI} bandwidth and GPU memory only to tensors that actually need them. By avoiding unnecessary migrations, \name{} minimizes thrashing and significantly improves runtime performance.

\subsection{Sensitivity}
\subsubsection{Sensitivity of Oversubscription Levels}
\begin{figure}[t]
    \centering
    \includegraphics[width=0.9\linewidth]{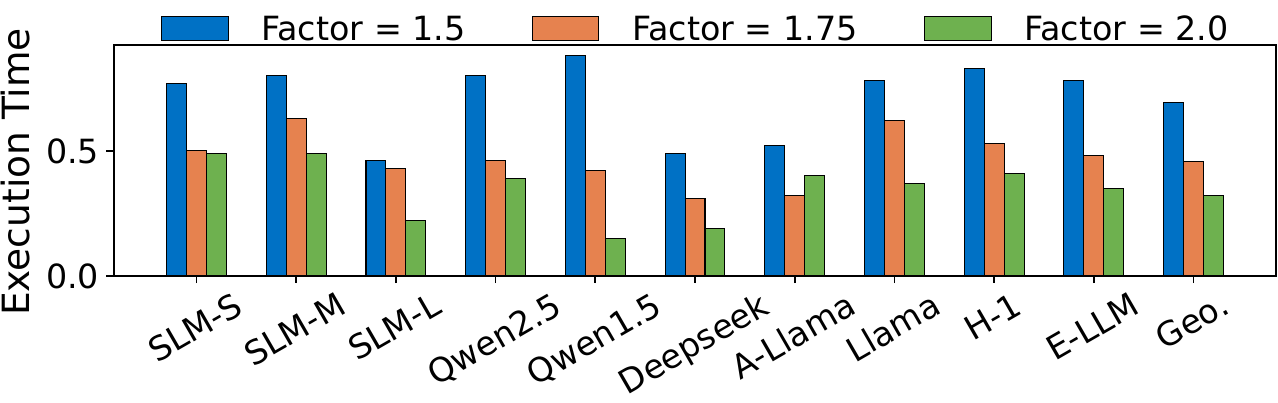}
    \caption{Execution time of \name{} normalized to Baseline under different oversubscription factor. }
    \vspace{-5pt}
    \label{fig:factor_comparison}
\end{figure}

We further evaluate the impact of different oversubscription factors. As shown in Figure~\ref{fig:factor_comparison}, \name{} achieves average speedups of $1.4\times$, $2.2\times$, and $3.1\times$ under oversubscription factors of $1.5$, $1.75$, and $2.0$, respectively. The gains increase as GPU memory becomes more constrained, indicating that \name{} is most effective when UVM paging is most severe. This trend underscores \name{}'s value as LLM memory demands grow and oversubscription becomes increasingly common.

\subsubsection{Sensitivity to GPU Platforms}
\begin{figure}[t]
    \centering
    \includegraphics[width=0.9\linewidth]{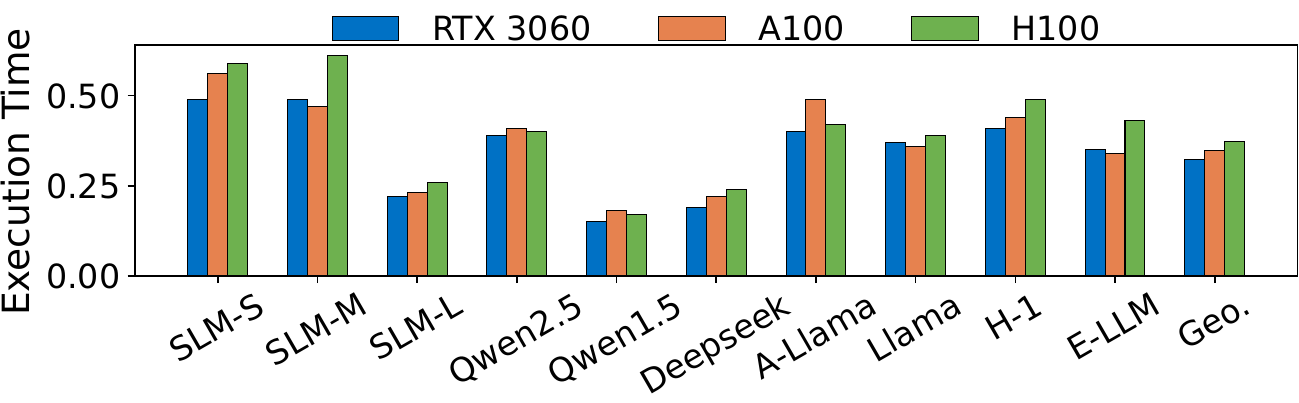}
    \caption{Execution time of \name{} normalized to Baseline under different GPU platform.}
    \label{fig:diff_gpus}
\end{figure}

Figure~\ref{fig:diff_gpus} evaluates \name{} across RTX 3060, A100, and H100 GPUs. \name{} consistently achieves substantial speedups of 3.1$\times$, 2.9$\times$, and 2.7$\times$, respectively. Although the gains decrease slightly on newer GPUs with higher interconnect bandwidth and larger memory capacity, \name{} remains effective across diverse GPU architectures.

\subsection{Ablation Study}
\begin{figure}[t]
    \centering
    \includegraphics[width=0.9\linewidth]{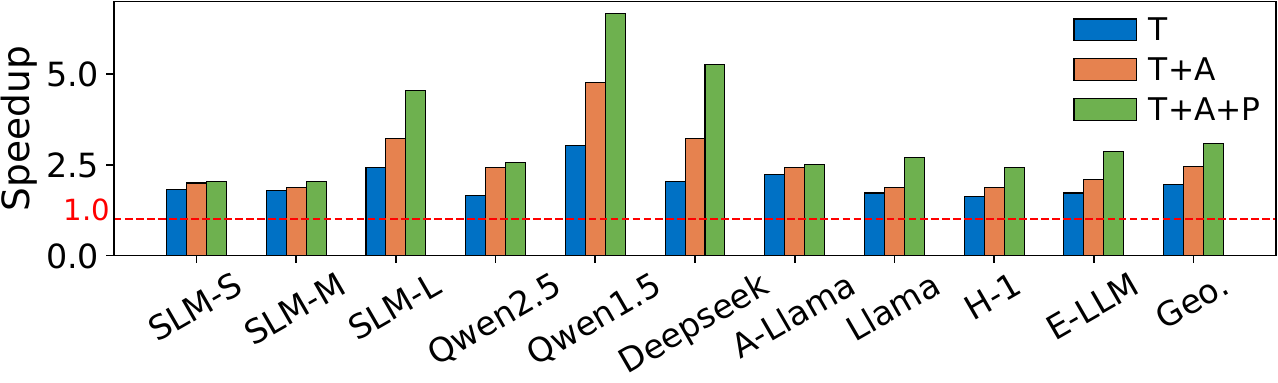}
    \caption{Ablation study of \name{}. 
\textit{T}: tensor-level prefetching; 
\textit{T+A}: adding look-ahead tensor affinity; 
\textit{T+A+P}: full design with policy-driven selectivity (higher is better). }
    \vspace{-5pt}
    \label{fig:ablation}
\end{figure}
To evaluate the contribution of each component, we compare three variants of \name{} against the baseline: \textbf{T}, which applies tensor-level prefetching; \textbf{T+A}, which adds look-ahead tensor affinity; and \textbf{T+A+P}, the full design with policy-driven selectivity. As shown in Figure~\ref{fig:ablation}, \textbf{T} delivers a $2.0\times$ speedup by avoiding unnecessary object-level migration. Adding affinity (\textbf{T+A}) increases the average speedup to $2.5\times$ by keeping highly reused tensors on the GPU. The full system (\textbf{T+A+P}) achieves the best results at $3.1\times$, showing that combining tensor-level visibility, affinity analysis, and policy-driven prefetching yields the highest performance.

\subsection{Energy and Power}
\begin{figure}[t]
    \centering
    \includegraphics[width=0.9\linewidth]{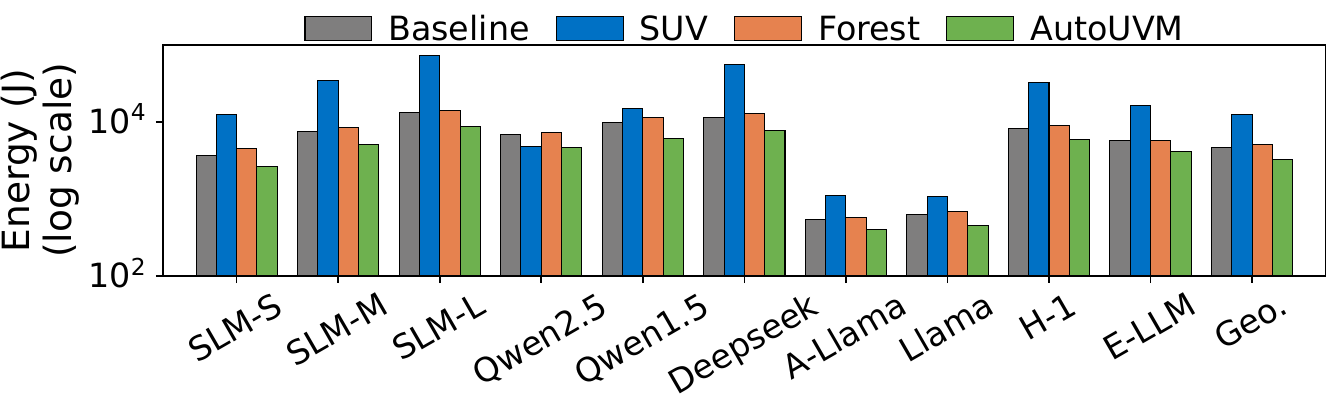}
    \caption{Total energy consumption across models (log scale).}
    \vspace{-5pt}
    \label{fig:energy}
\end{figure}

Figure~\ref{fig:energy} shows total energy consumption across all models (log scale). \name{} leaves average GPU power essentially unchanged (a 3--4W increase), as it does not alter kernels, GPU frequency, or computation. Yet it cuts total energy by 31\% on average over baseline UVM and up to 74\% over object-level prefetchers, as shorter execution time due to fewer page faults and unnecessary transfers outweighs the small power increase.

\begin{figure}[t]
\centering
\begin{minipage}{0.46\linewidth}
  \centering
  \includegraphics[width=\linewidth]{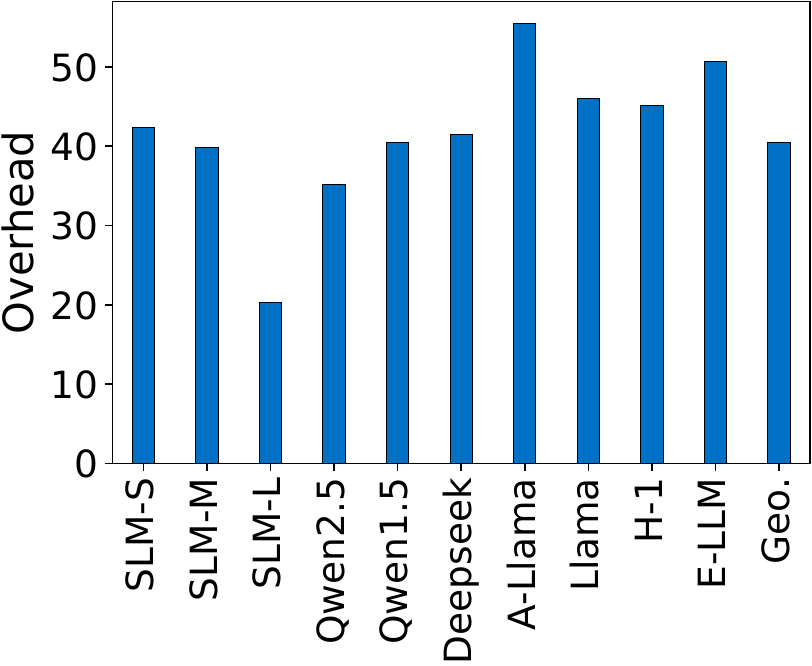}
  \caption{Profiling overhead across models.}
  \label{fig:overhead}
\end{minipage}
\hspace{0.01\linewidth}
\begin{minipage}{0.46\linewidth}
  \centering
  \includegraphics[width=\linewidth]{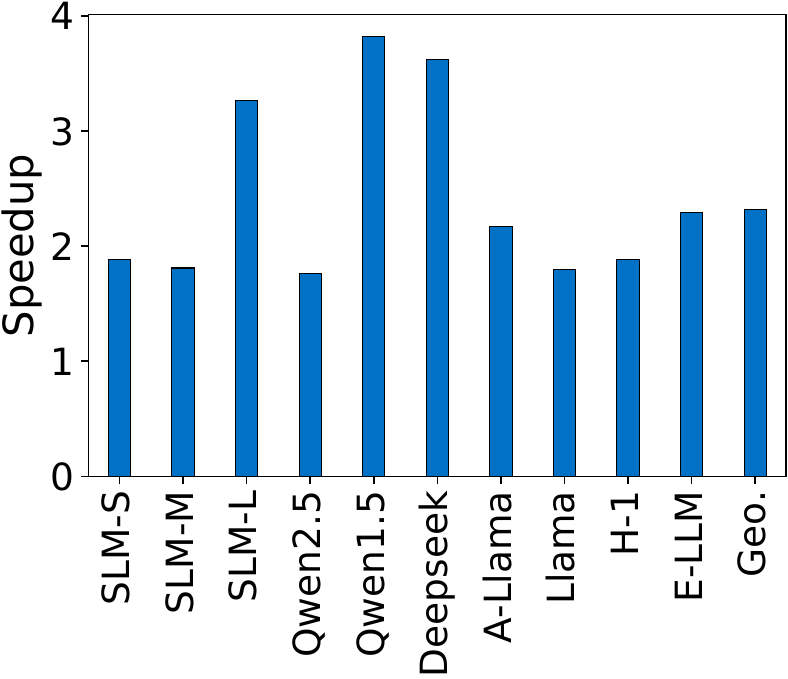}
  \caption{Performance with fused kernels.}
  \label{fig:fused_kernel_perf}
\end{minipage}
\vspace{-10pt}
\end{figure}

\subsection{Profiling Overhead}
Figure~\ref{fig:overhead} reports the cost of \name{}'s offline profiling pass, normalized to a single inference iteration ($41.5\times$ on average). This is a \textit{one-time, offline} cost: the profile is generated once and reused across all subsequent inference runs, so it is fully amortized and does not affect steady-state performance.

\subsection{Generalization to Fused Kernels}

To verify that \name{} generalizes beyond eager-mode execution, we evaluate it with kernel fusion enabled, where multiple operators are consolidated into single kernels with coarser tensor-access patterns. We apply \texttt{torch.compile} to all evaluated models, letting the PyTorch Inductor backend fuse elementwise, normalization, and attention operations, and regenerate \name{}'s offline profiles against the fused execution. As shown in Figure~\ref{fig:fused_kernel_perf}, \name{} achieves an average $2.3\times$ speedup over baseline UVM under an oversubscription factor of 2.0, confirming that its tensor-level prefetching remains effective under the more complex access patterns introduced by kernel fusion.

\section{Related Work}

\noindent\textbf{Profile-Guided and Runtime-Adaptive Prefetching.}
\label{sec:related-profiling}
DL workloads execute iteratively, enabling offline profiling to capture stable access patterns for later execution. Capuchin~\cite{peng2020capuchin} profiles one iteration to guide prefetching, eviction, and recomputation, while TERAIO~\cite{yuan2026cost} profiles tensor lifetimes to plan fine-grained offloading. Neither targets UVM prefetching. On the UVM path, Forest~\cite{lin2025forest} combines compile-time pattern detection with runtime access classification but relies on hardware access counters. These systems require source instrumentation or hardware support, and their runtime feedback adjusts prefetch aggressiveness rather than transfer necessity. In contrast, \name{} is software-only, decoupling offline kernel--tensor profiling from a lightweight online monitor and enabling tensor-level prefetching through transparent allocator hooks without model-code changes.

\noindent\textbf{UVM Prefetching.}
Prior work improves UVM through prefetching and data placement. SUV~\cite{suv2024} and Forest~\cite{lin2025forest} operate at the granularity of \texttt{cudaMallocManaged} allocations, while DeepUM~\cite{jung2023deepum} uses correlation-based prediction to prefetch future pages from historical fault patterns. However, these approaches lack framework-level tensor-management semantics exposed by modern DL frameworks. SUV and Forest further assume uniform access behavior within each managed object, using static~\cite{suv2024} or hardware-counter--driven~\cite{lin2025forest} access-pattern analysis. None consider the interaction between UVM and framework-level memory management.

Ganguly et al.~\cite{ganguly2019interplay} improve the UVM built-in prefetcher using access locality and intensity, while EarlyAdaptor~\cite{go2023early} adapts prefetching based on page-fault history. These methods target general workloads and are orthogonal to software UVM primitives such as \texttt{cudaMemPrefetchAsync} and \texttt{cudaMemAdvise}.
To the best of our knowledge, \name{} is the first automated prefetching system that incorporates modern DL-framework semantics to enable programmer-agnostic software prefetching.


\section{Conclusion}
We presented \name{}, an automated, framework-aware UVM prefetching system for oversubscribed LLM execution.
\name{} exposes fine-grained tensor behavior to UVM and migrates only performance-critical data, guided by a single offline profile and lightweight online runtime monitoring, without modifying model code.
Across 10 LLMs, \name{} achieves an average $3.1\times$ speedup over baseline UVM, outperforms the best-performing prior prefetcher (DeepUM) by $1.9\times$, and improves over object-level prefetchers by up to $4.7\times$, while reducing page faults and energy.
\name{} offers a practical, programmer-agnostic path for memory-intensive LLM workloads on commodity GPUs.


\section*{Acknowledgements}
This work was supported by NSF grants, 2541979, CCF-2452081, and CAREER-2341039.

\bibliographystyle{IEEEtran}
\bibliography{references}

\end{document}